\documentclass[aps,prl,twocolumn,groupedaddress,showpacs]{revtex4}
\usepackage{graphicx}

\begin{document}


\title{Crossovers in ScaleFree Networks on Geographical Space}

\author{Satoru Morita}
\email[]{morita@sys.eng.shizuoka.ac.jp}

\affiliation{Department of Systems Engineering,  
Shizuoka University, 3-5-1 Hamamatsu 432-8561, Japan}

\date{\today}

\begin{abstract}
Complex networks are characterized by several topological properties: 
degree distribution, clustering coefficient, average shortest path
length, etc.
Using a simple model to generate scale-free networks embedded
on geographical space,
we analyze the relationship between 
topological properties of the network and    
attributes (fitness and location) of the vertices
in the network. 
We find there are two crossovers for varying the 
scaling exponent of the fitness distribution.
\end{abstract}

\pacs{89.75.Hc; 89.75.Da}

\maketitle

Many natural, social, and technological
systems can be described in term of complex networks,
in which vertices represent interacting units,
and edges stand for interactions among them
\cite{strogatz,albert02,dorogovtsev,newman}.
The complex networks, which are far from absolutely regular or complete
random, are characterized by several properties:
degree distribution, clustering coefficient $C$, 
average shortest path length $L$, etc.
Many real networks exhibit 
a scale-free degree distribution $P(k)\sim k^{-\nu}$,
typically with scaling exponent $2<\nu<3$
\cite{strogatz,albert02,dorogovtsev,newman,barabasi,amaral}.
Most real networks have a large clustering coefficient,
which is defined as the
probability that a pair of vertices with a common
neighbor are also connected to each other \cite{watts98}.
The local clustering coefficient usually decreases with 
the degree \cite{ravasz}.
In addition, small-world effect is seen 
in many networks \cite{watts98}. 
For many network models,
the average shortest path length grows
logarithmically $L\propto \ln N$ or more slowly
\cite{dorogovtsev,newman,cohen}.

In order to understand the structure of
the complex networks,
many models have been proposed.
Such models can be grouped into two main classes.
The first class contains growing networks
with preferential attachment.
Barab\'{a}si and Albert (BA) proposed this type model originally 
\cite{barabasi}.
For BA model, the scaling exponent $\nu$ is always 3
and its clustering  coefficient is relatively small for large size.
Then, several modified models have been presented
to reproduce the realistic aspects of networks
\cite{dorogovtsev, ravasz, krapivsky, bianconi}. 
The second class contains static networks,
where each vertex has a intrinsic fitness
measuring the its importance or rank
\cite{caldarelli,boguna,soderberg,servedio,masuda}.
In several models, the location in geographical space
is also taken into consideration 
\cite{rozenfeld,yook,dall,barthelemy,warren,manna,yang,masuda2,andrade}.
In such models, the topological properties of the network are 
essentially determined by the characteristics of the vertices.
The purpose of this paper is 
to make clear the relationship between
the topology of the network and the attributes of the vertices
for a simple model of the second class.
We find when the scaling exponent $\gamma$ of the fitness distribution 
varies, there are two crossovers at $\gamma=2$ and $\gamma=3$.

Our model is defined in the following.
We consider $N$ vertices.
We assume that each vertex has an fitness $a_i (i=1,2,\dots,N)$.
For simplicity, the fitness values are assigned deterministically as
\begin{equation}
 a_i=\left(\frac{i}{N}\right)^{\frac{1}{1-\gamma}} \ (i=1,2,\dots,N).
\end{equation}
for $\gamma>1$
The case of $\gamma\simeq 2$ is known as Zipf's law.
When $N$ is adequately large, 
the distribution of the fitness is given approximately as
\begin{equation}
 \rho(a) = (\gamma-1)a^{-\gamma}
+\frac{\delta(a-1)}{2N}+\frac{\delta(a-N^{\frac{1}{\gamma-1}})}{2N}.
\label{eq_rho}
\end{equation}
in the finite support
\begin{equation}
1 \leq a \leq N^{\frac{1}{\gamma-1}}.
\label{eq_kukan}
\end{equation}
Here $\delta(x)$ denotes Dirac's delta function.
Thus, the distribution of the fitness 
follows the power law $\rho(a)\propto a^{-\gamma}$ 
with slight adjustments at the both side.
In addition, the vertices are distribute
randomly in $d$-dimensional space with 
uniform distribution.
For simplicity, the distance is defined by L-max norm, and
the boundary condition is periodic.
We assume that the fitness and the location
are independent mutually. 
The condition to link vertices $i$ and $j$ is 
\begin{equation}
\frac{(2 l(i,j))^d}{a_i a_j}<\theta ,
\label{eq_cond}
\end{equation}
where $l(i,j)$ denotes the distance
between these vertices and $\theta$ is a threshold.
Here, the threshold value $\theta$ is chosen so that the
total number of connections equals $mN$.
Thus the average degree is given by
$ \langle k \rangle= 2m$.
The network resulting from our method has
a scale-free degree distribution as is shown in Fig.~1(a).

Since the vertices follow the uniform distribution in the unit 
$d$-dimensional cube,
the probability to link a pair of vertices
with fitness $a$ and $a'$
is given as
\begin{equation}
r(a,a')=\min(\theta a a',1) .
\label{eq_raa}
\end{equation}
The average degree for a vertex with fitness $a$ 
is calculated as
\begin{equation}
\bar{k}(a)  =   \displaystyle
N\int r(a,a') \rho(a')da' .
\label{eq_ka}
\end{equation}
Inserting (\ref{eq_rho}) and (\ref{eq_raa}) into (\ref{eq_ka}),
we obtain the approximate form for large $N$
\begin{equation}
\bar{k}(a) \simeq\left\{
\begin{array}{ll}
\displaystyle
\frac{2(\gamma-1)N-\gamma N^{\frac{1}{\gamma-1}}}
{2(\gamma-2)}\theta a
& (a<\theta^{-1}N^{\frac{-1}{\gamma-1}})\\
\displaystyle
\frac{(\gamma-1)\theta a-(\theta a)^{\gamma-1}}{\gamma-2}N
& (a \geq \theta^{-1}N^{\frac{-1}{\gamma-1}})\\
\end{array}\right. \ .
\label{eq_barka}
\end{equation}
We can estimate the threshold value $\theta$
from the fact that the average degree is described as  
\begin{equation}
2m =
\displaystyle 
\int \bar{k}(a) \rho(a)da  \ .
\label{eq_mN}
\end{equation}
Inserting (\ref{eq_rho}) and (\ref{eq_barka}) into (\ref{eq_mN}),
we get for large $N$
\begin{widetext}
\begin{equation}
2m\simeq N \frac{2(\gamma-1)^2\theta
-(3\gamma^2-5\gamma+2)\theta N^{\frac{2 - \gamma}{\gamma-1}}+ 
(\gamma^2-\gamma) \theta^{\gamma-1} - 
4(\gamma-2)\theta^{\gamma-1}\ln N - 
2(\gamma^2 - 3\gamma + 2) \,\theta^{\gamma-1}\,
\ln \theta}{2(\gamma -2)^2}
\end{equation}
\end{widetext}
The asymptotical solution $\theta$ for large $N$ is described as
\begin{equation}
\theta \simeq \left\{
\begin{array}{ll}
\displaystyle
\frac{2m}{N}\left(\frac{\gamma-2}{\gamma-1}\right)^2
& (\gamma>2)\\ 
\displaystyle
\left(\frac{2m(2-\gamma)}{N \ln N }\right)^{\frac{1}{\gamma-1}}
& (1<\gamma<2).
\end{array} \right.
\label{eq_sc}
\end{equation}
Thus, the asymptotical behavior changes at $\gamma=2$.
For $\gamma>2$, the asymptotic form of Eq.~(\ref{eq_barka})
is given as
\begin{equation}
\bar{k}(a)\simeq
2m\frac{\gamma-2}{\gamma-1}a \hspace{5mm} (\gamma>2).
\label{eq_ka2}
\end{equation}
Accordingly, $\bar{k}(a)$
is proportional to $a$.
On the other hand, for $1<\gamma < 2$, 
Eq.~(\ref{eq_barka}) is approximately 
\begin{equation}
\bar{k}(a)\simeq \frac{2m}{\ln N}a^{\gamma-1} \hspace{5mm} (1<\gamma < 2)
\label{eq_ka1}
\end{equation}
for $a>\theta^{-1}N^{\frac{-1}{\gamma-1}}$.
Thus, $\bar{k}(a)$ follows a power law decay with exponent $\gamma-1$.

\begin{figure}[tb]
\includegraphics[scale=0.48]{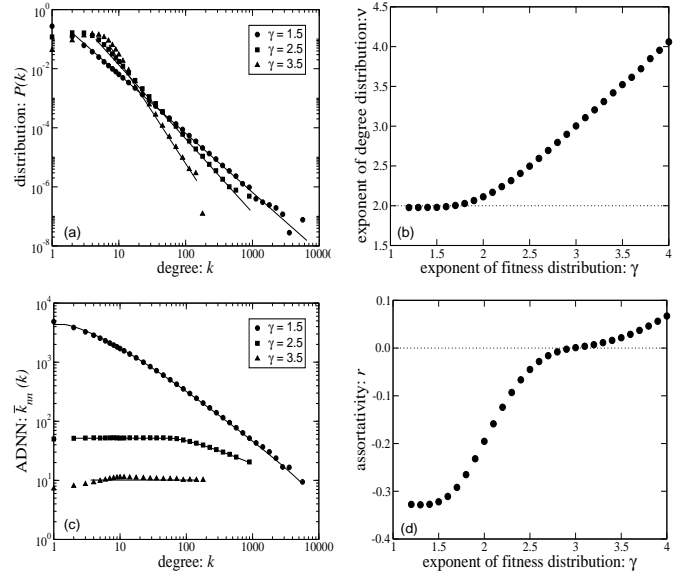}
\caption{(a) The degree distribution $P(k)$  obtained numerically
for $m=3$, $d=2$, $N=10000$, and $ \gamma = 1.5$ (circles),
$ \gamma = 2.5$ (squares)  
and $\gamma = 3.5$ (triangles).
This data is averaged over 100 realizations, and
the bin is taken logarithmically to reduce noise.
The solid curves stands for
the theoretical prediction (\ref{eq_distri1}) and (\ref{eq_distri2}).
(b) The exponent values 
calculated using the maximum likelihood method are shown 
as a function of $\gamma$ for $m=3$, $d=2$, $N=10000$.
(c) The average nearest neighbor degree (ANND)
for the same parameters as in (a).
The solid curves correspond to theoretical results (\ref{eq_annd1}),
(\ref{eq_annd2}), and (\ref{eq_annd3}).
(d) The assortativity $r$ (degree correlation) 
obtained numerically is 
shown as a function of $\gamma$ for the same parameters as in (b).
\label{fig_1}}
\end{figure}

Let us now calculate the degree distribution.
The degree distribution is calculated as
\begin{equation}
 P(k)  =  \int P(k|a) \rho(a)da .
\label{eq_degree}
\end{equation}
Here, the conditional probability $P(k|a)$ that vertex with 
fitness $a$ has degree $k$ is given by binominal form:
\begin{equation}
P(k|a)=\left(\begin{array}{c} N \\ k\end{array} \right) 
\left(\frac{\bar{k}(a)}{N}\right)^k 
\left(1-\frac{\bar{k}(a)}{N}\right)^{N-k} .
\end{equation}
For $\gamma>2$,
if the inside of the integral of (\ref{eq_degree}) has the maximum  
in the range (\ref{eq_kukan}), the integral is
approximated by using the gamma function.
Accordingly, in the region 
\begin{equation}
2m\frac{\gamma-2}{\gamma-1}+\gamma<k
<2m\frac{\gamma-2}{\gamma-1}N^{\frac{1}{\gamma-1}} +\gamma,
\label{eq_hanni}
\end{equation}
the degree distribution (\ref{eq_degree}) is described as
\begin{equation}
 P(k)\simeq 
\frac{(2m)^{\gamma-1}}{N^{\gamma-1}}
\frac{(\gamma-2)^{\gamma-1}}{(\gamma-1)^{\gamma-2}}
\frac{N!}{k!} \frac{\Gamma(k-\gamma+1)}{\Gamma(N-\gamma+2)} .
\label{eq_distri}
\end{equation}
For $k\gg 1$, we get the scale-free degree distribution:
\begin{equation}
 P(k)\simeq (2m)^{\gamma-1}
\frac{(\gamma-2)^{\gamma-1}}{(\gamma-1)^{\gamma-2}}k^{-\gamma} 
 \hspace{5mm} (\gamma >2).
\label{eq_distri1}
\end{equation}
In this case, the scaling exponent equals that of the fitness distribution.
For $\gamma < 2$, 
if the inside of this integral of Eq.~(\ref{eq_degree})
has the maximum in the range 
$(\theta^{-1}N^{\frac{-1}{\gamma-1}},N^{\frac{1}{\gamma-1}})$,
the degree distribution is calculated 
in the same way to the case of $\gamma>2$.
Accordingly, in the region 
\begin{equation}
\frac{5-2\gamma}{2-\gamma}<k
<\frac{2mN}{\ln N}+2,
\label{eq_hanni2}
\end{equation}
the degree distribution is described as
\begin{equation}
 P(k)\simeq \frac{2m}{\ln N} k^{-2}
 \hspace{5mm} (1< \gamma < 2).
\label{eq_distri2}
\end{equation}
As a result, in this case, the degree distribution 
is independent of $\gamma$, and 
the exponent is always 2.
These analyses are consistent with the numerical results 
(see Fig.~1 (a) and (b)).
Note the degree distribution is independent of the 
dimension $d$ in the both cases.

In addition to the degree distribution,
we study the degree-degree correlation $P(k'|k)$,
which measures the probability of a vertex with degree
$k$ to be linked to a vertex with degree $k'$.
In order to characterize this correlation,
it is useful to work with the average nearest 
neighbor degree (ANND), which is 
defined as $\bar{k}_{nn}(k)\equiv \sum_{k'} k' P(k'|k)$
\cite{satorras}.
Before estimating $\bar{k}_{nn}(k)$,
we estimate the ANND of a vertex with fitness $a$,
which is calculated as
\begin{equation}
\bar{k}_{nn}(a)=\frac{\displaystyle \int r(a,a')\bar{k}(a')\rho(a')da'}
{\displaystyle \int r(a,a')\rho(a')da'}+1 \ .
\label{eq_knna}
\end{equation}
Here the last term adding one is due to the fact that
the nearest neighbor vertex has at least one connection.
Eliminating $a$ with using (\ref{eq_ka2}),
we obtain an approximation for $\bar{k}_{nn}(k)$.
For $\gamma>3$, we obtain the asymptotical form for large $N$
\begin{equation}
\bar{k}_{nn}(k)\simeq
\frac{2m(\gamma-2)^2}{(\gamma-1)(\gamma-3)}+1 \hspace{5mm} 
(\gamma>3) .
\label{eq_annd1}
\end{equation}
Thus, the ANND $\bar{k}_{nn}(k)$ is independent of $k$.
This result indicates there is no correlation between degrees of linked pairs.
However the numerical result shows
there is a small positive correlation (see Fig.~1 (c)).
This correlation may be due to the fluctuation 
of the vertex density in the $d$-dimensional space.
For $2<\gamma<3$, the asymptotical form of ANND is 
\begin{equation}
\bar{k}_{nn}(k)\simeq\left\{
\begin{array}{l}
\displaystyle
A\left[\frac{\gamma+1}{2(\gamma-1)}N^{\frac{3-\gamma}{\gamma-1}}
-1\right]+1\\
\hspace{15mm} (k<\frac{\gamma-1}{\gamma-2}N^{\frac{\gamma-2}{\gamma-1}},
 2<\gamma<3)\\
\displaystyle
A\left[\alpha\frac{N^{3-\gamma}}{k^{3-\gamma}}-
\beta
\frac{N^{\frac{1}{\gamma-1}}}
{k}-1\right]+1\\
\hspace{15mm} (k>\frac{\gamma-1}{\gamma-2}N^{\frac{\gamma-2}{\gamma-1}},
 2<\gamma<3)
\end{array}\right. \ ,
\label{eq_annd2}
\end{equation}
where $A=\frac{2m(\gamma-2)^2}{(\gamma-1)(3-\gamma)}$,
 $\alpha=\frac{(\gamma-1)^{3-\gamma}}{(\gamma-2)^{4-\gamma}}$, and 
$\beta=\frac{\gamma(3-\gamma)}{2(\gamma-2)^2}$.
This result indicates that
the ANND is constant for small $k$ and 
decays approximately $\bar{k}_{nn}(k)\propto k^{-(3-\gamma)}$ for large $k$. 
For $1<\gamma<2$, we obtain
\begin{equation}
\bar{k}_{nn}(k)=\left\{
\begin{array}{l}
\displaystyle
\frac{2m N(2-\gamma)(2\gamma-1)}{\gamma \ln N} +1\\
\hspace{15mm} 
(k<1/(2-\gamma), 1<\gamma<2)\\
\displaystyle
4 m N\frac{\ln k+\ln (2-\gamma)+\gamma-1/2}
{(2k-1)\ln N }+1\\
\hspace{15mm}
(k>1/(2-\gamma), 1<\gamma<2)
\end{array}\right. \ .
\label{eq_annd3}
\end{equation}
This result indicates that
the ANND decays $\bar{k}_{nn}(k)\propto \frac{\ln k}{k}$ for large $k$. 
Figure 1 (c) shows these analyses agree well with 
the numerical results.
In addition, we calculate numerically
the assortativity $r$ defined by Newman \cite{newman2}
for several values of $\gamma$.
The assortativity denotes
Pearson correlation coefficient of the degrees at either 
ends of an edge.
Figure 1 (d) shows
the network has positive (negative) degree correlation 
if $\gamma>3$ ($\gamma<3$).

\begin{figure}[tb]
\includegraphics[scale=0.48]{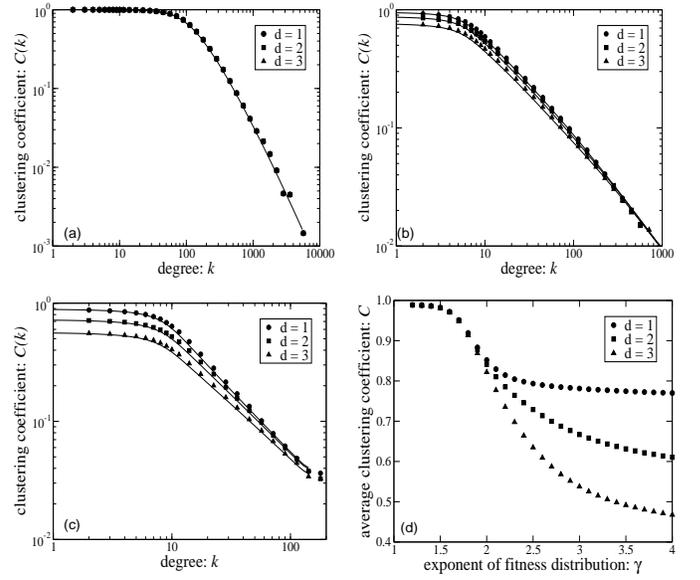}
\caption{
The clustering coefficient as a function of $k$
for $m=3$, $N=10000$, and
$\gamma=1.5$ (a), $\gamma=2.5$ (b), and $\gamma=3.5$ (c),
where circles, squares and triangles
correspond to the numerical results for $d=1$, $d=2$, and $d=3$,
respectively. 
In (a), these three plots coincide exactly.
The solid curves correspond to the 
predictions from the numerical integrate of (\ref{eq_ca}).
In (d), the average clustering coefficient as a function of 
$\gamma$ for $m=3$, $N=10000$ and
$d=1$ (circles), $d=2$ (squares), and  $d=3$ (triagles).
\label{fig_2}}
\end{figure}

The clustering coefficient is calculated as
a function of $a$ as follows 
\begin{equation}
C(a)=\frac{\displaystyle\int r_3(a,a',a'')\rho(a')\rho(a'') da' da''}
{\displaystyle\int r(a,a')r(a,a'')\rho(a')\rho(a'') da' da''}
\label{eq_ca}
\end{equation}
where $r_3(a,a',a'')$ denotes
the probability that 
three vertices with fitness values $a$, $a'$, and  $a''$
form a triad.
Since it is difficult to analytically calculate
the numerator of (\ref{eq_ca}),
we resort to numerical integration.
We can calculate also $C(k)$ by the numerical integration of
the product of $C(a)$ and $P(a|k)$ which is given by Bayes's law 
$P(a|k)=P(k|a)\rho(a)/P(k)$ (see Fig.~2).
For $\gamma>2$ the clustering coefficient $C(k)$ decreases
with the dimension $d$.
On the other hand, for $\gamma<2$, the 
clustering coefficient $C(k)$ seems to be
independent of the dimension $d$.
This suggests that when $\gamma<2$,
the spatial structure is irrelevant to the network structure.
This fact is confirmed by 
the behavior of the average cluster coefficient defined by Watts and 
Strogatz \cite{watts98}, as is shown in Fig.~2 (d).

Finally we study 
the average shortest path length $L$ (Fig.~3).
For $\gamma>3$, 
the average shortest path length seems to
follow a power law $L\propto N^\mu$,
where $\mu$ is somewhat smaller than $1/d$.
On the other hand,
for $\gamma<3$,
the average shortest path length grows more slowly than $\ln N$.
In this case, 
a pair of vertices with sufficiently large fitness are
always linked, 
because $\max(\theta a_i a_j)=\theta N^{\frac{2}{\gamma}-1} \gg 1$
for large $N$.
Consequently, some vertices, which 
connect each other regardless of their distance,
compose a shortcut network.
As a result, the spatial structure is irrelevant to 
the average shortest path length,
and thus the network is ultrasmall \cite{cohen}.

\begin{figure}[t]
\includegraphics[scale=0.48]{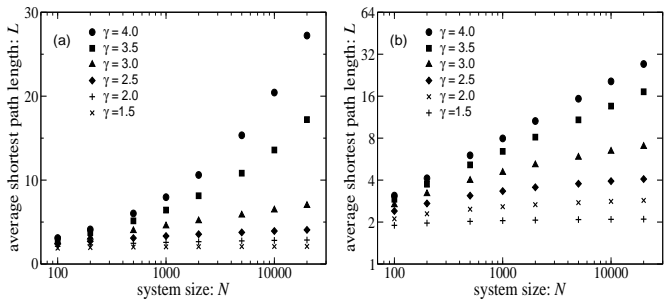}
\caption{
(a) Log-linear plot of 
the average shortest path length $L$ vs. the number 
of vertices $N$
for different values of $\gamma$ ($\gamma=4.0, 3.5, 3.0, 2.5, 2.0, 1.5$
from top to bottom) and $d=2$.
(b) The same data in log-log plot.
\label{fig_3}}
\end{figure}

In summary,
we have studied a scale free network embedded on geographical space.
While the scaling exponent of the 
degree distribution equals
 $\gamma$ for the scaling exponent $\gamma$ of the fitness distribution
for $\gamma>2$,
it is always 2 for $\gamma<2$.
For $\gamma<2$, the spatial effect is irrelevant
to some topological properties (ANND or clustering coefficient) 
of the network.
While the network is disassortative (negative degree correlation) 
for $\gamma<3$,
it is weakly assortative (positive degree correlation) for $\gamma>3$.
Moreover, the network is not small for $\gamma>3$,
whereas the spatial effect is irrelevant
to the average shortest path length for $\gamma<3$.
Thus, there are two crossovers at $\gamma=2$ and $\gamma=3$.
The reason why the crossovers are observed clearly
is that the fitness is assigned deterministically.
If we use random fitness following the power law,
the crossovers became somewhat blurred,
but do not change qualitatively. 
Furthermore, the
preliminary numerical research suggests that
these results hold even if the distribution of the
location of the vertices is not uniform.
Therefore, we expect that
the results presented in this paper are robust
in the condition that the nearer pairs tend to be linked.

\begin{acknowledgments}
This research was carried out under the ISM Cooperative 
Research Program 2006-ISM CRP-1008.

\end{acknowledgments}


\begin{thebibliography}{99}

\bibitem{strogatz}
S. H. Strogatz,
Nature {\bf 410}, 268 (2001).

\bibitem{albert02}
R. Albert and A. -L. Barab\'{a}si,
Rev. Mod. Phys. {\bf 74}, 47 (2002).

\bibitem{dorogovtsev}
S. N. Dorogovtsev and J. F. F. Mendes,
Adv. Phys. {\bf 51}, 1079 (2002);
{\it Evolustion of Networks} (Oxford University Press, New York, 2003).

\bibitem{newman}
M. E. J. Newman,
SIAM Review {\bf 45}, 167 (2003).

\bibitem{barabasi}
A. -L. Barab\'{a}si and R. Albert,
Science {\bf 286}, 509 (1999)

\bibitem{amaral}
L. A. N. Amaral, A. Scala, M. Barth\'{e}lemy, and H. E. Stanley,
Proc. Nat. Acad. Sc. USA {\bf 97}, 11149 (2000).

\bibitem{watts98}
D. J. Watts and S. H. Strogatz,
Nature {\bf 393}, 440 (1998).

\bibitem{ravasz}
E. Ravasz, A. L. Somera, D. A. Mongru, Z. Oltvai,
and  A.-L. Barab\'{a}si,
Science {\bf 297}, 1551 (2002);
E. Ravasz, A.-L. Barab\'{a}si, 
Phys. Rev. E 67, 026112 (2003).

\bibitem{cohen}
R. Cohen and S. Havlin,
Phys. Rev. Lett. {\bf 90}, 058701 (2003).

\bibitem{krapivsky}
P. L. Krapivsky, S. Redner, and F. Leyvraz,
Phys. Rev. Lett. {\bf 85}, 4629?4632 (2000).

\bibitem{bianconi}
G. Bianconi and A.-L. Barab\'asi, 
Phys. Rev. Lett. {\bf 86}, 5632 (2001).

\bibitem{caldarelli}
G. Caldarelli, A. Capocci, P. De Los Rios, and M. A. Mu\~{n}oz
Phys. Rev. Lett. {\bf 89}, 258702 (2002).

\bibitem{boguna}
M. Bogu\~{n}a and R. Pastor-Satorras, 
Phys. Rev. E {\bf 68}, 036112 (2003).

\bibitem{soderberg}
B. Soderberg, Phys. Rev. E {\bf 66}, 066121 (2002).

\bibitem{servedio}
V. D. P. Servedio, G. Caldarelli, and P. Butta,
Phys. Rev. E {\bf 70}, 056126 (2004).

\bibitem{masuda}
N. Masuda, H. Miwa, N. Konno,
Phys. Rev. E {\bf 70}, 036124 (2004).

\bibitem{rozenfeld}
A. F. Rozendeld, R. Cohen, D. ben-Avraham, and S. Havlin,
Phys. Rev. Lett. {\bf 89}, 218701 (2002);
D. ben-Avraham, A. F. Rozenfeld, R. Cohen, and S. Havlin,
Physica A {\bf 330}, 107 (2003). 


\bibitem{andrade}
J. S. Andrade, Jr., H. J. Herrmann, R. F. S. Andrade, and
L. R. da Silva,
Phys. Rev. Lett. {\bf 94}, 018702 (2005).


\bibitem{yook}
S.-H. Yook, H. Jeong, and A.-L. Barab\'{a}si, 
Proc. Natl. Acad. Sci. U.S.A. {\bf 99}, 13 (2002)

\bibitem{dall}
J. Dall and M. Christensen,
Phys. Rev. E {\bf 66}, 016121 (2002).

\bibitem{barthelemy}
M. Barth\'elemy
Europhys. Lett. {\bf 63}, 915 (2003).

\bibitem{warren}
C. P. Warren, L. M. Sander, and I. M. Sokolov,
Phys. Rev. E {\bf 66}, 056105 (2002).

\bibitem{manna}
S. S. Manna and P. Sen,
Phys. Rev. E {\bf 66}, 066114 (2002); {\bf 66}, 066114 (2002).

\bibitem{yang}
K. Yang, L. Huang, and L. Yang,
Phys. Rev. E {\bf 70}, 015102(R) (2004).

\bibitem{masuda2}
N. Masuda, H. Miwa, N. Konno,
Phys. Rev. E {\bf 71},  036108 (2005).

\bibitem{satorras}
R. Pastor-Satorras, A. Vazquez, and A. Vespignani,
Phys. Rev. Lett. {\bf 87}, 258701 (2001).

\bibitem{newman2}
M. E. J. Newman,
Phys. Rev. Lett. {\bf 89}, 208701 (2002).










\end{thebibliography}

\end{document}